\documentclass[article, eqsecnum]{revtex4}
\usepackage{graphicx,latexsym,amsmath}

\def\beq{\begin{equation}}
\def\eeq{\end{equation}}
\def\bea{\begin{eqnarray}}
\def\eea{\end{eqnarray}}
\def\alinv{\left(\frac{1}{\alpha}\right)}

\def\d{\partial}

\newcommand{\deriv}[2]{\frac{\partial #1}{\partial #2} }

\begin{document}

\title{Charged Black Hole in a Canonical Ensemble}

\author{Andrew P. Lundgren}
\affiliation{Department of Physics, Cornell University, Ithaca, New York 14853}
\email{apl27@cornell.edu}

\begin{abstract}
We consider the thermodynamics of a charged black hole enclosed in a cavity.  The charge in the cavity and the temperature at the walls are fixed so that we have a canonical ensemble.  We derive the phase structure and stability of black hole equilibrium states.  We compare our results to that of other work which uses asymptotically anti-deSitter boundary conditions to define the thermodynamics.  The thermodynamic properties have extensive similarities which suggest that the idea of AdS holography is more dependent on the existence of the boundary than on the exact details of asymptotically AdS metrics.
\end{abstract}

\maketitle

\section{Introduction}

A black hole in asymptotically flat space is thermodynamically unstable when the temperature at infinity is fixed.  To solve this problem we can place the black hole inside a finite spherical cavity.  The temperature is fixed at the surface of the cavity, which could be physically realized by placing a heat bath around the cavity.  The black hole can now be thermodynamically stable, which is partly due to the fact that the horizon of the black hole can be near the point where the temperature is fixed.  In this paper we are specifically interested in the thermodynamic ensemble where the cavity also contains a fixed amount of charge; this is an example of a canonical ensemble.  The grand canonical ensemble, where the electric potential is fixed rather than the charge, was considered in \cite{BBWY}, and we will use many of the methods developed there.

Many papers use the alternate method \cite{hawkingpage} of adding a negative cosmological constant to put the black hole in anti-de Sitter (AdS) space.   The properties of AdS space stabilize the black hole by acting as a reflecting box.  There has been much interest in the thermodynamics of anti-de Sitter space because of the conjectured AdS holography \cite{ADSCFT, witten, thooft}.  It appears that there is a duality between the thermodynamics of an AdS black hole and a field theory in one fewer dimension.  It is not clear whether the duality is a result specifically of the properties of anti-de Sitter space, or whether it is simply a result of the confinement provided by the reflecting walls.

We are therefore motivated to compare the results of our analysis with the results obtained using the AdS method, in particular the work by Chamblin et.al. \cite{CEJM1,CEJM2}.  They compared the AdS charged black hole with the liquid-gas transition in classical thermodynamics, and demonstrated that the influence of the conserved electric charge yields a distinctive phase structure.  This paper shows that the same phase structure exists when the black hole is in a finite box with no cosmological constant, which is a strong suggestion that the AdS holography is a result of confinement rather than the properties of AdS space.  As we were finishing this paper, we discovered a paper by Carlip and Vaidya \cite{carlip} with the same motivation.   That work focused on the critical exponent of the second-order phase transition; we will focus more on the detailed phase structure.

The cavity we are considering has surface area $4 \pi r_B^2$ and temperature $T$.  The simplest case is when the charge inside the cavity is fixed to be zero.  Then there is a minimum temperature $k T = \frac{3 \sqrt{3} \hbar}{8 \pi r_B}$ below which no black hole can be in equilibrium inside the cavity.  Above this temperature, there are two states of the black hole that are possible.  The larger one is quite close in size to the cavity; the cavity is within the 3M photon orbit of the black hole, where $2M$ is the Schwarzschild radius.  This state is locally but not globally stable, and given enough time it will tunnel to the state with only radiation inside the cavity.  The other state which is possible at the same temperature is a smaller, unstable black hole which corresponds to a maximum of the free energy.  Therefore, it also sets the height of the barrier that prevents tunneling to the state with only radiation (hot flat space).

Above $k T = \frac{27 \hbar}{32 \pi r_B}$, the larger radius solution becomes globally stable, and now hot flat space can ``decay'' to form a black hole.  There is a still an unstable state of smaller radius that controls the height of the barrier between the stable states.  As the temperature rises, the height of the barrier decreases, and the radius of the unstable black hole shrinks.  At the same time, the size of the stable black hole solution is increasing until at infinite temperature it merges with the surface of the cavity.

In the canonical ensemble the charge of the black hole is fixed and the only variable allowed to fluctuate is the radius of the horizon.  With zero charge, there were no equilibrium states other than flat space at low temperature, but there were two black hole states in addition to flat space at higher temperature.  The structure becomes more complicated for non-zero charge.  For charges less than $(\sqrt{5} - 2) r_B$ in natural units, there is a range of temperatures for which there are three possible black hole states.  The largest and smallest are locally stable, and the one between must be unstable.  For values of the temperature outside the range, or for any temperature when the charge is above the critical value, there is only one solution, which is always at least locally stable.  One special case is when the temperature is very low, and so the black hole should have a very small radius.  The danger is that the inner and outer horizons can merge, and yield an extremal black hole or a naked singularity (the extremal case is the dividing line between having a horizon or not).  For any finite temperature, this does not happen, and the horizons are always separated.

We will begin by discussing the physical situation we are considering and the boundary conditions necessary to implement the canonical ensemble.  Then we derive the action and use it to find the possible equilibria given a certain temperature and charge.  The uncharged and grand canonical ensembles are reviewed before looking in detail at the canonical ensemble.  We find the number and stability of equilibria over the entire temperature - charge phase space, and compare to the AdS result.  Some slices of the free energy function are displayed which are identical in structure to those of the AdS result.  We end with a discussion of the meaning of these similarities and some directions for future research.

\section{The Geometry and Action}
We start with the usual form of the static spherically symmetric spacetime.  The only free functions are $b$ and $\alpha$, which are both functions only of $r$.  We analytically continue the time coordinate by defining $\tau = i t$ to give a positive-definite metric.  The metric takes the form
\beq
ds^2 = b^2 d\tau^2 + \alpha^2 dr^2 + r^2 ( d\theta^2 + \sin^2 \theta \; d\phi^2) \label{metric}
\eeq
which is the same as in \cite{BBWY}, except simplified slightly.  Derivatives with respect to $r$ will be denoted by primes.  Throughout this paper, we will refer to a value of $r$ as the radius, although the physical meaning of $r$ is not the distance to the origin but instead gives the area of a sphere concentric with the origin.  This is a quantity that can be measured without leaving the sphere, and is not dependent on the function $\alpha$.  Also we use natural units $k = c = G = \hbar = \frac{1}{4 \pi \epsilon_0} = 1$, where mass, charge, and inverse temperature have units of distance.

The ``Euclideanized'' metric forces the $\tau$ coordinate to be periodic because $b(r)$ shrinks to zero at the horizon and the $t - r$ plane behaves like the center of a disk there.  The periodic imaginary time coordinate thermalizes the space (see \cite{gibhawk, yorkeuclid} for details).  For convenience,  we choose the period of $\tau$ to be $2 \pi$.  In the Euclideanized action, inverse temperature at some radius is the period in the $\tau$ direction.  The inverse temperature at a radius $r$ is $2 \pi b(r)$, which depends on $r$ as a result of the gravitational redshift. 

We define the thermodynamic ensemble by fixing the temperature at the outer boundary, and also fixing the electric field which serves to fix the charge inside.  The coordinate position of the boundary will be $r = r_B$ and the area is $4 \pi r_B^2$.  The outer boundary is the full three-dimensional metric on the hypersurface $r = r_B$; we need the value of $b$ to completely specify this metric.  In terms of the reciprocal temperature $\beta = T^{-1}$, the boundary condition is
\beq
\int b(r_B) d\tau = 2 \pi b(r_B) = \beta ~.
\eeq
Note that $b$ is analogous to the lapse $N$, which in the Schwarzschild metric is $\sqrt{1-2M/r}$.  Approaching the horizon, this quantity decreases toward zero.
 
The inner boundary of the system is at the event horizon of the Euclidean black hole.  The position will be denoted $r = r_+$.  Since this is the horizon, we must have
\beq
b(r_+) = 0
\eeq
and since the $\tau - r$ part of the metric looks like a disc, we must avoid a conical singularity by requiring
\beq
\left. (\alpha^{-1} b') \right|_{\, r_+} = 1 ~. \label{regularity}
\eeq
 
The gravitational action is
\beq
I_g = -\frac{1}{16 \pi} \int_M d^4 x \sqrt{g} R + \frac{1}{8 \pi} \int_{\d M} d^3 x \sqrt{\gamma} ( K - K_0 )
\eeq
which is the usual Hilbert action with the metric fixed on the spatial boundary of the system $\d M$.  $K$ is the extrinsic curvature of the timelike boundary of the system, which is the surface $r = {\rm constant}$.  $K_0$ is a subtraction term so that the action for flat space will be zero.  It is calculated by embedding the same surface in flat spacetime.  The results are
\bea
\sqrt{-g} R &=& - \frac{2 b}{\alpha} + 2 b \alpha - \frac{4 r b~'}{\alpha} - 4 r b \alinv ' - 2 r^{2} \left( \frac{b~'}{\alpha} \right)' \\
K &=& - \frac{1}{\alpha b r^2} (b r^2)' \quad ; \qquad K_0 = - \frac{2}{r} ~.
\eea
The $K$ and $K_0$ terms are only present at the outer boundary, because that is where the metric is fixed.  The quantites that are fixed are $r$ and $b$; $\alpha$ is free to vary because it involves the direction normal to the boundary.  We need to integrate by parts any terms with second derivatives of $b$ or first derivatives of $\alpha$.  The integration by parts in the bulk term yields total derivatives that cancel the boundary term involving $K$ while depositing some terms on the inner boundary.  The action, after integrating over angles and performing the integration by parts, is
\beq
I_g = - \pi \int_{r_+}^{r_B} \left( \left( \frac{2 r}{\alpha} \right) b' + \left( \alpha + \frac{1}{\alpha} \right) b \right) dr
+ \left. 2 \pi b r \frac{{}}{{}} \right|_{r_B}
- \left. 2 \pi \frac{r b}{\alpha} \right|_{r_+}
- \left. \pi r^2 \frac{b'}{\alpha} \right|_{r_+} ~.
\eeq
Of the three terms after the integral, the first is the remnant of the $K_0$ subtraction which makes the energy zero when $r_+ = 0$ which corresponds to flat space.  The middle term is zero because $b(r_+) = 0$.  Using the regularity condition \eqref{regularity} the last term becomes one-fourth the horizon area, which is the standard result for the entropy of a black hole.  The entropy term is a direct result of the periodicity of the $\tau$ coordinate.

The electromagnetic action is very simple because we are only interested in a spherically-symmetric static electric field, so the only potential we need is $A_\tau (r)$.  The variation of $A_\tau$ in the Maxwell action gives the curved space version of Gauss's Law.  As in \cite{BBWY}, we can use this and the analytic continuation to a Euclidean metric to show that the charge $e$ in the cavity is defined by
\beq
\frac{r^2}{b \alpha} A_\tau' = - i e ~. \label{fixcharge}
\eeq
The Maxwell action simplifies to (integrating over angles and $\tau$)
\beq
I_{EM} = \pi \int dr \left( \frac{r^2}{\alpha b} A_\tau' \right) A_\tau' - \left. 2 \pi \left( \frac{r^2}{\alpha b} A_\tau' \right) A_\tau \right|_{r_B}
\eeq
where the second term is included so that we are fixing the quantity \eqref{fixcharge} on the boundary rather than the potential.  On the inner boundary, we have another regularity condition.  The potential in an orthonormal basis is $A_{\hat \tau} = b^{-1} A_\tau$.  At the inner boundary, $b$ goes to zero so we also fix $A_\tau = 0$ there so that the physical potential does not become unbounded.  The form of the action is already suitable for fixing the potential on the inner boundary so we can now proceed to deriving the thermodynamics from the action.

We vary the action to obtain the equations of motion for $b$, $\alpha$, and $A_\tau$.  The solutions depend on $r_+$ and $e$, and give the usual metric of a charged black hole.  The reduced action $I_*$ is defined as the value of the action evaluated for the solution we have obtained, which is
\beq
I_*(r_+; \beta, e) = \beta r_B \left( 1 - \sqrt{\left( 1 - \frac{r_+}{r_B} \right) \left( 1 - \frac{e^2}{r_+ r_B} \right)} \right) - \pi r_+^2 ~.
\eeq
The last term is the entropy of the black hole, as was mentioned earlier.  The first term is $\beta$ times the quasilocal energy of the black hole \cite{brownyork} evaluated as the surface of the cavity.  This gives the nice result that the free energy (which is $I_* / \beta$) is $E - T S$ with the quasilocal energy playing the role of $E$.

\section{Equilibria and Stability}

It is convenient to define non-dimensionalized variables using the radius of the boundary as a standard length:
\beq
I \equiv \frac{I_*}{4 \pi r_B^2} , \quad x \equiv \frac{r_+}{r_B} , \quad q \equiv \frac{e}{r_B} , \quad \sigma \equiv \frac{\beta}{4 \pi r_B} , \quad \Theta \equiv 4 \pi r_B T ~.
\eeq
Now the conditions for a physical solution are easy to write.  We must have $0 < x < 1$, and $0 < q < x$, so that the horizon exists but is smaller than the outer boundary and larger than the inner horizon.  The charged black hole solution has two horizons; when the Schwarzschild radius is equal to the charge the horizons merge to give an unphysical extremal Reissner-Nordstr\"om solution.

The action written in these new variables is
\beq
I(x; \sigma, q) = \sigma \left( 1 - \sqrt{(1 - x)(1 - \frac{q^2}{x})} \right) - \frac{x^2}{4} ~.
\eeq
Note that the physical action scales linearly with the area of the boundary.

The stationary points of the action will dominate the path integral.  These stationary points are the equilibria of the system at the given temperature and charge.  The only free variable is $x$, so stationary points are defined by
\beq
\deriv{I}{x} = \frac{1}{2} \left[\frac{\sigma (x^2 - q^2)}{x^2 \sqrt{(1 - x)(1 - \frac{q^2}{x})}} - x \right] = 0 ~.\label{didx}
\eeq
The stability of an equilibrium point is determined by the curvature of the action.  A negative second derivative indicates that the equlibrium is unstable.  Small fluctuations of $x$ will tend to grow; if the black hole absorbs a small amount of excess heat it will grow and the amount of energy it radiates to the boundary will decrease.  The black hole will continue to absorb energy and grow until it reaches a stable equilibrium or one of the edge cases where $x = 1$ or $x = q$.

A local minimum of the action is a stable equilibrium, for which small fluctuations do not grow.  States like this may only be metastable if there are other states with lower action.  There are two possibilities.  The state with lowest action may also be an equilibrium, which we will then call globally stable.  It is also possible that one of the edge cases has the lowest action, which most likely means that the the actual globally stable equilibrium is not a member of the ensemble we have defined.  We will see that in the canonical action this does not happen.  Since we are only considering a subset of the possible metrics and neglecting the action of the radiation that should be present in the cavity, there may still be some other state with lower action.

We can make a single graph that shows the solutions and their stability at once.  First we introduce the free energy
\beq
F = (E - T S) = I / (\beta) .
\eeq
The condition for a solution is that the derivative should be zero.  This means that the temperature of a solution is
\beq
T(x_s) = \frac{E'}{S'}
\eeq
which is certainly no surprise.  Now, treating the temperature as a function of $x$, we have
\beq
T' = \frac{E''}{S'} - \frac{E' S'}{(S')^2} .
\eeq
and examining the second derivative of F yields
\beq
F'' = E'' - T S'' = S' T' .
\eeq
$T$ is actually fixed, so $T'$ should not be thought of as the derivative of the actual temperature, but is merely the slope of the temperature graph.  However, since $S'$ is always positive in this case, we can simply plot the temperature as a function of $x$ and determine the stability by looking at the graph.

\section{Uncharged Case}
Setting $q = 0$ and solving for the zeros of \eqref{didx}, we obtain the cubic equation 
\beq
x^3 - x^2 + \sigma^2 = 0 .
\eeq
We can solve this for $\sigma$ and substitute back into the equations for the action and its second derivative to determine the local and global stability of solutions.  If we let $x_e$ denote a value of $x$ that is an equilibrium, we have:
\bea
\sigma = x_e \sqrt{1 - x_e} \\ 
I_s = x_e (\sqrt{1 - x_e} - 1 + \frac{3}{4} x_e) \\
\deriv{^2I}{x^2} = \frac{(3 x_e - 2)}{4 (1 - x_e)}
\eea

The lowest temperature (largest sigma) for which the action has a local minimum is $\sigma = \frac{2 \sqrt{3}}{9}$.  The size of the black hole is $x_s = \frac{2}{3}$ and so the cavity that encloses the black hole is located at the $3M$ photon orbit.  A local minimum means that the black hole will be at least metastable.  The solution will be globally stable if the action is a global minimum.  There are no other local minima, but we need to check the action at the extreme values of $x$, which are $I(x=0) = 0$ and $I(x=1) = \sigma - \frac{1}{4}$.  The solution we are considering has positive action, although it is less than that of the $x = 1$ solution.  As a result, the black hole is stable, but eventually it will tunnel to hot flat space.  It is clear that there must be a state with maximum action between the local minimum and the $x = 0$ solution.  This state is an unstable black hole with a smaller radius than the stable solution at the same temperature.

For global stability, the action has to be negative, which occurs at $x = \frac{8}{9}$, and $\sigma = \frac{8}{27}$.  As the temperature is increased from this point, the radius of the black hole increases, but the action remains less than the action at the outer boundary ($x = 1$).  At any temperature above this point, flat space is unstable and can tunnel to the black hole solution.  There is always another solution 

\section{Charged Case}
The Reissner-Nordstr\"om metric of a charged black hole is
\beq
ds^2 = - \left( 1 - \frac{2M}{r} + \frac{e^2}{r^2} \right) dt^2 + \left( 1 - \frac{2M}{r} + \frac{e^2}{r^2} \right)^{-1} dr^2 + r^2 d\Omega^2
\eeq
with Lorentzian signature; with the $\tau = i t$ substitution this is the metric that we would find from our equations of motion.  The main new feature is the existence of two horizons, at $r_\pm = M \pm \sqrt{M^2 - e^2}$.  It is convenient to make $r_+$ and $e$ the independent variables because $r_+$ determines the entropy and $e$ determines the electromagnetic field.  If the charge grows large enough that $r_+ = e$, or $x = q$ in rescaled variables, the two horizons merge.  This is the extremal Reissner-Nordstr\"om solution; if the charge is increased any more, the horizon disappears and the singularity is visible to the external universe.  This super-extremal Reissner-Nordstr\"om solution is not desirable from a physical standpoint, and we will show that the electric potential required to reach this solution is unrealistically high.

In the grand canonical ensemble, the equations giving a solution are:
\bea
& & (1 - \phi^2) x^3 - x^2 + (1 - \phi^2)^2 \sigma^2 = 0 \\
& & q = \frac{\phi \, x^2}{\sigma (1 - \phi^2)} .
\eea
We now introduce a trick for solving these equations, which can also be somewhat useful in the canonical ensemble.  We look for a solution with a given ratio of $q$ and $\sigma$ to $x$.  Define
\beq
q \equiv \epsilon x , \quad \sigma \equiv b x .
\eeq
The result is
\beq
x = 1 - \frac{1 - \epsilon^2}{\epsilon^2} \frac{\phi^2}{1 - \phi^2}
\eeq
In the GCE, $q$ only stands for the expectation value of the charge, and the potential $\phi$ is the quantity that is fixed.  This result shows that when the expectation value of the charge coincides with the event horizon, then both must also coincide with the boundary, which means this is not a good physical solution.  For $\phi < 1$, the non-extremal solutions have $x < 1$; they are physical and super-extremal solutions are not.  The situation reverses for $\phi > 1$, but this corresponds to about $10^{27}$ volts (the Planck voltage).  So for any reasonable values of the potential, the black hole is sub-extremal.

\section{Canonical Ensemble}
Turning to the canonical ensemble, the charge is fixed so we only vary $x$.  We have only one equation to consider, which simply states that at a solution, denoted $x_s$, the action should be a local extremum.  The condition for an equilibrium is therefore
\beq
\frac{\sigma (x_e^2 - q^2)}{x_e^2 \sqrt{(1 - x_e)(1 - \frac{q^2}{x_e})}} - x_e = 0 \label{ceeq} ~.
\eeq

We start with the high-temperature behavior where $\sigma$ goes to zero.  The equilibria in that case are determined from $x_e^2 (1 - x_e) (x_e - q^2) = 0$, which gives $x = 0, q^2, 1$.  Of these, only $x = 1$ is a physically meaningful solution, and it indicates that the event horizon will tend to merge with the outer boundary at extremely high temperature.

Now we turn to the low-temperature behavior, where $\sigma \gg 1$.  For simplicity, assume that $q$ is positive.  The first term in \eqref{ceeq} will dominate, so the only solutions are near $x = q$.  The second term must be negative, so actually we have $x > q$.  This shows that the solution stays non-extremal for any non-vanishing temperature.  Both the canonical and grand canonical ensembles will avoid the extremal case given physically reasonable conditions.  Since the charge is fixed in this ensemble, we might have worried that as the temperature is lowered, the black hole will shrink without losing any charge, and the horizons will merge to give a naked singularity.  However, the conserved quantity tends to improve the stablity of the ensemble, and seems to prevent the black hole from decaying to a smaller radius.

At slightly higher temperatures, the entropy of the black hole makes more of a contribution to the free energy.  It is now possible for larger black holes to form, because the energy that they have to gain from the heat bath is accompanied by an increase in the entropy.  In fact, it is possible for the interplay between these effects to produce multiple solutions for a given temperature and charge.  If there are multiple solutions, we must consider which ones are locally stable or unstable.

The first case is when the charge is larger than a critical value.  For any temperature there is a single solution, and it is always stable.  The size of the black hole grows with the temperature, and it does not grow to the size of the outer boundary for any finite temperature.  The solution is globally stable relative to the states we are considering, because it has lower free energy than the endpoints.  It is possible that there are other states with the same charge and temperature but perhaps a different topology or non-spherical metric fluctuations.  If these states have a lower free energy, than the charged black hole solution would be only metastable and could decay to them.

\begin{figure}
\tiny
\includegraphics[width=2.5in]{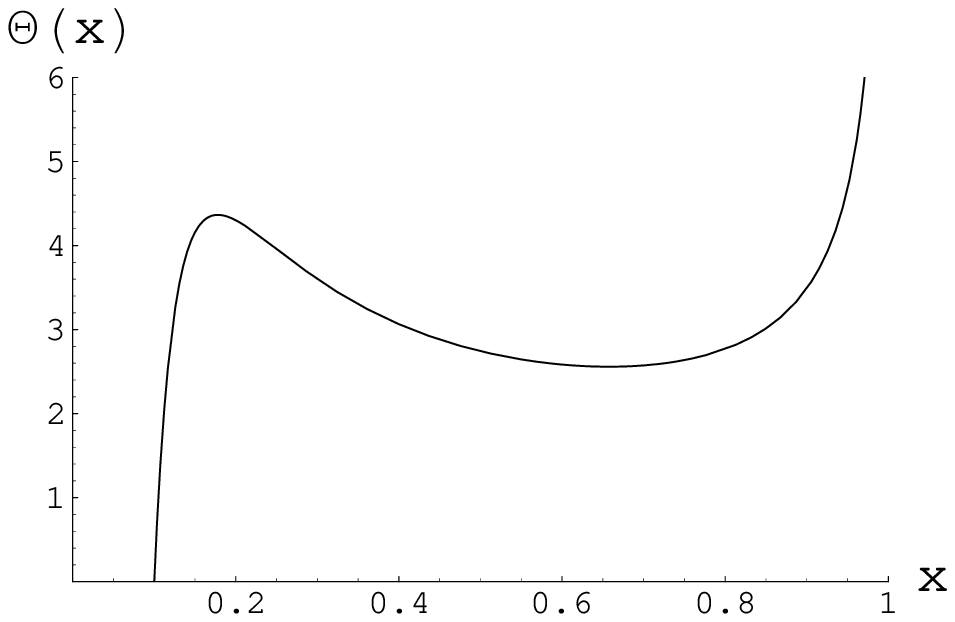}                                          
\includegraphics[width=2.5in]{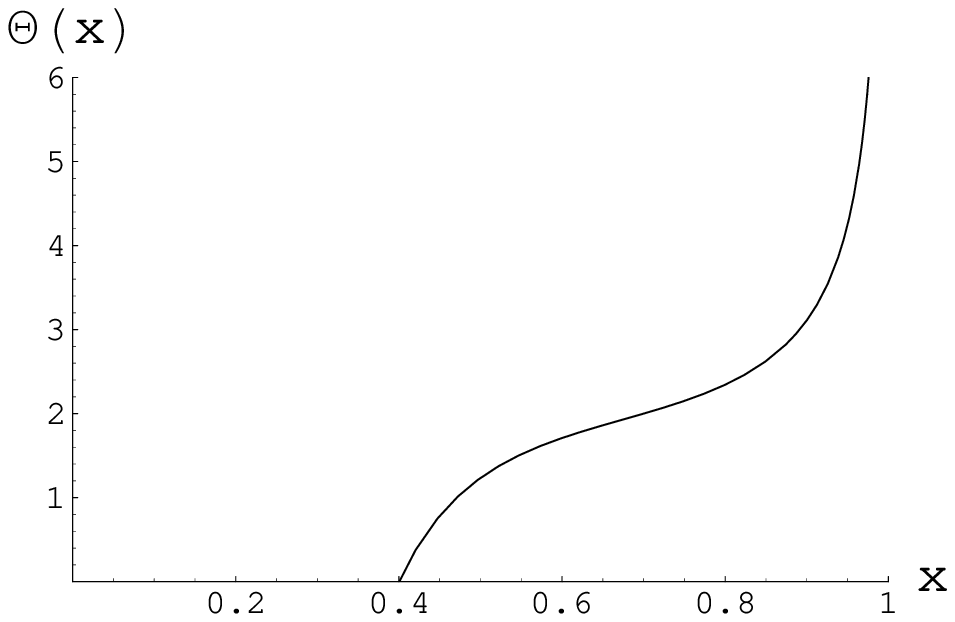}                                          
\caption{Temperature versus x for q = 0.1 and q = 0.4.}
\label{actionfig}
\end{figure}

When the charge drops below the critical value, there exists a band of temperatures for which three states can coexist.  At a given temperature, the smallest and largest radius equilibria are stable, and since the temperature graph must change slope between them, the intermediate radius solution is always unstable.  Whichever solution has the lower free energy will be globally stable, because once again the endpoints have higher free energy.  The intermediate unstable solution is a local maximum and so it sets the height of the barrier between the two stable states.  As we approach the critical charge, the barrier becomes lower and the three states approach the same radius until they merge into one solution which must be marginally stable.

Outside the band of temperatures where there are three solutions, the behavior is the same as for larger charge.  The low temperatures have nearly extremal black holes and the high temperatures have very large ones, and both are stable.

\begin{figure}
\tiny
\includegraphics[width=2.5in]{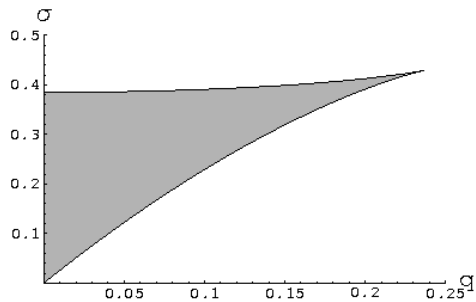}                                          
\caption{Phase diagram with charge $q$ on the horizontal axis and temperature $T$ on the vertical.  The shaded region has three equilibria as opposed to only one outside.}
\label{phasedgmfig}
\end{figure}

The temperature - charge phase space is divided into the region with one solution and the region with three coexisting solutions.  The dividing line is formed by the states that are marginally stable, that is, the second derivative of the free energy is zero.  The transition from three solutions happens because the unstable solution merges with one of the stable solutions and annihilates it, leaving only a single solution.  The dividing line can be found by solving for the position where the temperature graph has zero slope.  It is easiest to find the charge $q$ as a function of $x$ for which $\d \Theta / \d x = 0$ which yields
\beq
q_{transition} = \left[ \frac{x}{6 x - 5} \left( x^2 + 3 x - 3 + \sqrt{( x - 9 ) ( x - 1 )^3} \right) \right]^{1/2} ~.
\eeq
In Figure \ref{phasedgmfig} the two regions in the $q - \Theta$ plane are plotted.  The critical charge above which there is never more one solution is $q_{crit} = \sqrt{5} - 2$.

We can now address the issue of global stability which is simpler here than in the uncharged case.  The most stable configuration could be any of the local minima, or one of the endpoints of the range where the action is defined.  When $x = q$ the horizons merge and we have a different topology, and when $x = 1$ the horizon merges with the boundary of the system.  In either case the action is no longer defined.  However, one of the stable equilibria must be the global minimum for all values of temperature and charge.  Throughout the phase space, there are always either two minima of the action with a maximum between them, or a unique minimum.  The continuity of the action then guarantees that one of the equilibria will always be the global minimum of the action, over the range $q < x < 1$ where the action is considered to be physically meaningful.  There are likely to be other states outside those which we have considered in our ensemble, particularly those where a gas of charged particles has replaced the black hole.  We will discuss this further in the conclusion.

\begin{figure}[h]
\tiny
\includegraphics[width=2.3in]{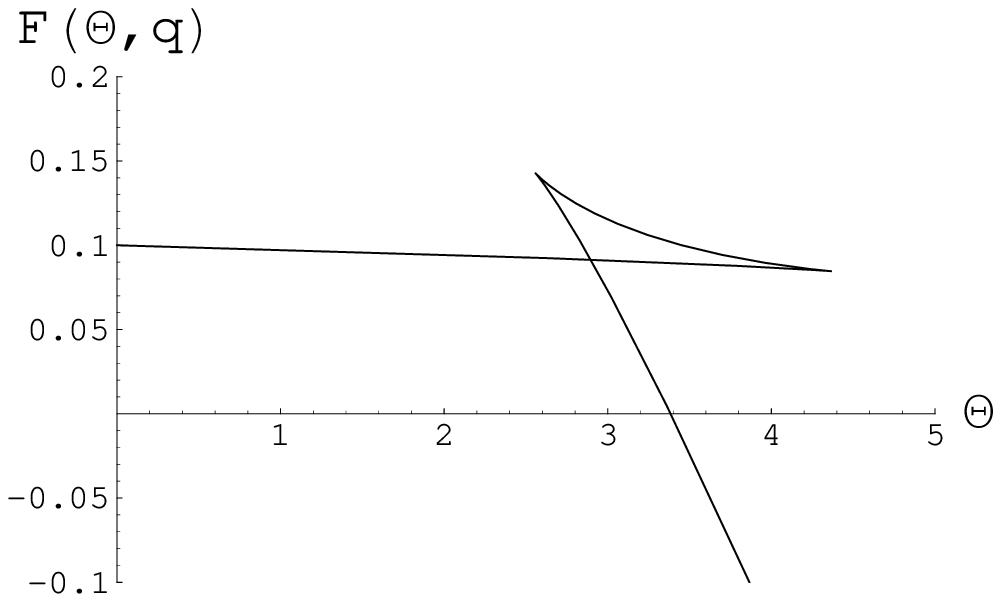}                                          
\includegraphics[width=2.3in]{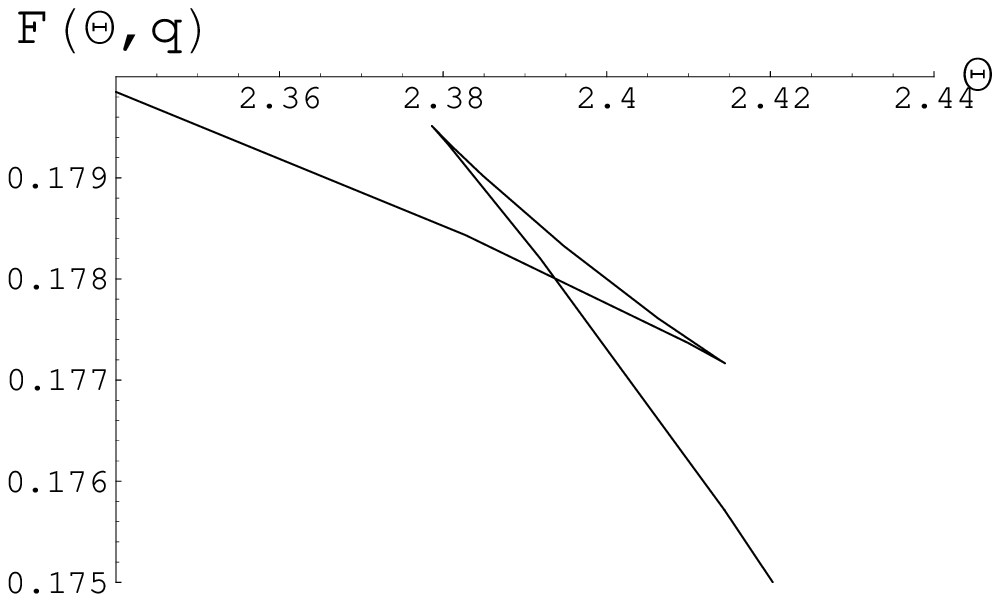}                                          
\includegraphics[width=2.3in]{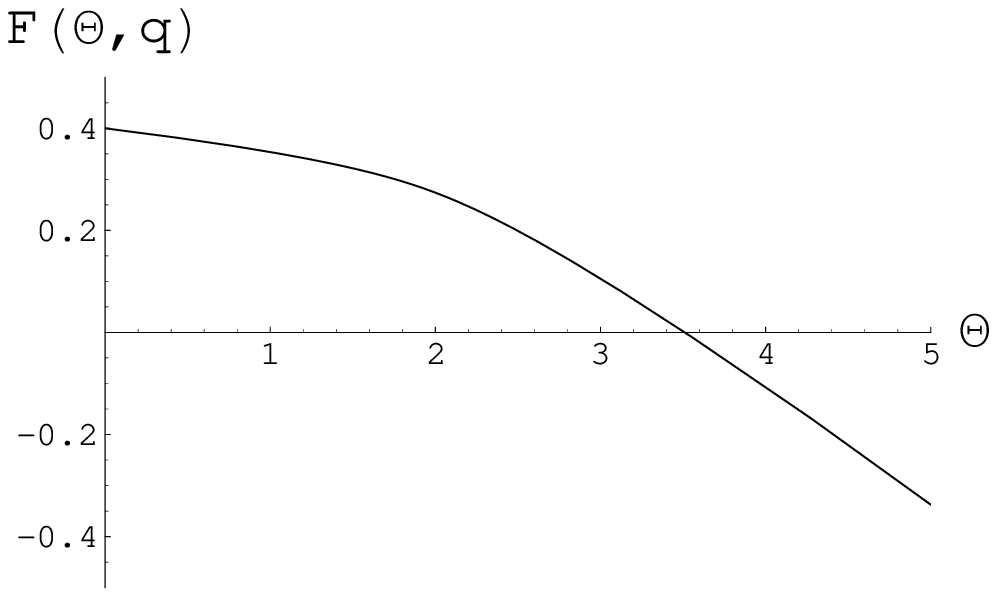}                                          
\caption{Free energy versus temperature for $q = 0.1$, $q = 0.22 \approx q_{crit}$, and $q = 0.4$.}
\label{fixchargefig}
\end{figure}

To make the phase structure more clear, we can plot the free energy $F$ against the temperature with the charge fixed.  When there are three equlibria, this will allow us to see which one is globally stable.  Figure \ref{fixchargefig} shows that above the critical charge, there is only one branch of the free energy.  As we raise the temperature, the black hole smoothly grows from a small near-extremal state to a large black hole the nearly fills the cavity.  Below the critical charge, the smooth transition from small to large black holes is interrupted in the band of intermediate temperatures where there are three equilibria.  As the temperature increases in this band, the free energy of the small black hole increases while that of the large one decreases.  They cross each other and the larger black hole becomes the globally stable one.  As the temperature increases past this point, the smaller black hole is still locally stable, so it could persist for a long time.  However, the free energy of this state and the unstable state are growing closer in value, so as the temperature increases there is less of a barrier preventing a transition to the globally stable state.  At some point, the branch of the free energy that we are on merges with the branch for the unstable state, and ends.  There is now only the larger black hole state, and this will continue to grow and be the only possible equilibrium as the temperature increases to infinity.

At the temperature where the free energies of the two stable equilibria are equal, the two phases (small black hole and large black hole) can coexist.  This is a slightly dangerous viewpoint, because we are only considering a single black hole concentric with the cavity; nothing in the analysis shows that two different-size black holes could coexist.  We can think of the black hole of being in a superposition of the two states, but they are macroscopic and of very different masses so interactions with the environment would cause decoherence.  The best viewpoint to take is that the two states are equally probable.

\begin{figure}[h]
\tiny
\includegraphics[width=2.3in]{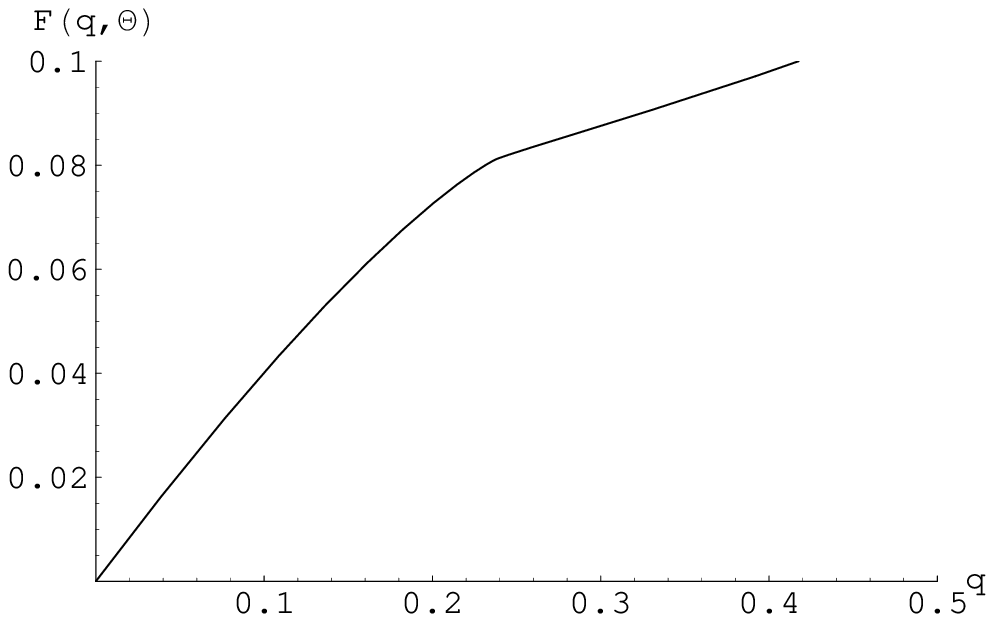}                                          
\includegraphics[width=2.3in]{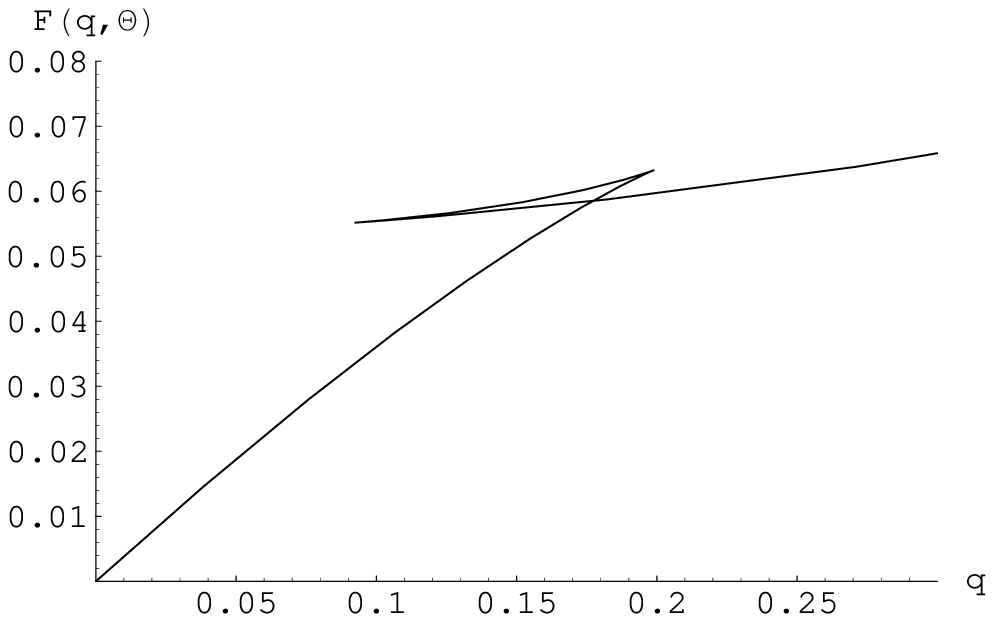}                                          
\includegraphics[width=2.3in]{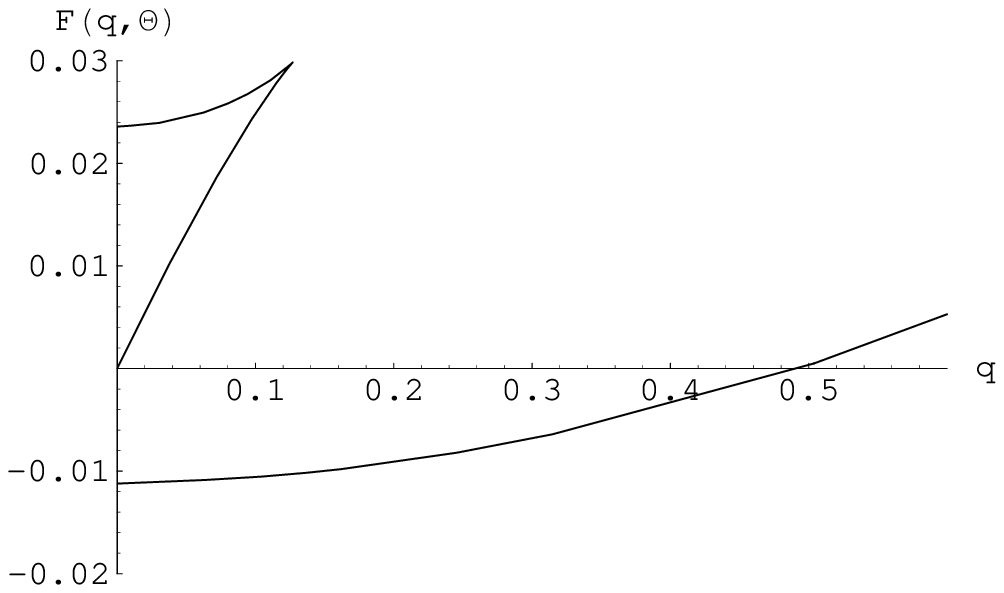}                                          
\caption{Free energy versus charge for $\theta = 2.33$, $\theta = 2.56$, and $\theta = 3.57$.}
\label{fixtempfig}
\end{figure}

Figure \ref{fixtempfig} graphs the free energy versus the charge for fixed values of the temperature.  The shapes, but not the values, can be compared with Figure 5 in \cite{CEJM2}.  Figure \ref{fixchargefig} can be compared with their Figure 4.  These are simply a more visual way of demonstrating that the detailed phase structure of the AdS charged black hole and the charged black hole in a cavity are essentially identical.

\section{Conclusion}
We have derived the thermodynamics of a charged black hole in a finite spherical cavity.  The charge governs the number of possible equilibria.  Above the critical value $q = (\sqrt{5} - 2) r_B$, there is always a single equilibrium state, which is a small black hole with a Schwarzschild radius only slightly larger than its charge in natural units.  As the temperature increases the size of the black hole increases, but for any finite temperature above zero, we always have physically reasonable behavior; the horizon never disappears as in the case of an extremal Reissner-Nordstr\"om solution nor does the horizon grow to touch the walls of the enclosing cavity.

For charges below the critical value, there is a band of temperatures where there are three different equilibria.  Figure \ref{phasedgmfig} shows the region of the temperature-charge phase space where this is the case.  One of the equilibria is always unstable and so it is not directly relevant to the thermodynamics.  The behavior in this temperature band is like a phase transition, from the smaller black hole state to the larger one.  At the critical temperature and charge, there is a second-order phase transition beyond which there is only one phase (one stable equilibrium rather than two stable and one unstable).  In \cite{CEJM2}, the critical exponent was derived for this transition in the AdS case, and in \cite{carlip} it was shown to be the same for the finite cavity.  The current paper has shown in more detail that the phase structure is the same.  It seems likely, therefore, that AdS holography does not depend on specific properties of AdS space but instead simply on the confinement of a reflecting box or cavity at a fixed temperature.

We have not discussed the issue of hot flat space in this paper, or the possibility of other topologies that may contribute to the ensemble.  When we showed that one of the equilibria is always a global minimum of the action, that only counts the states that we considered in deriving the action.  More exotic topologies could possibly have a smaller value of the action, which would allow our ``globally stable'' state to decay to the more exotic topology.

A more important and possibly more tractable concern is the Hawking-Page \cite{hawkingpage} transition to hot flat space.  With the charge fixed to zero, this is likely to happen because flat space has a lower value of the action than the black hole unless the temperature is high enough.  We have the problem that there is no solution for flat space with a nonzero charge.  In reality, the black hole can emit charged Hawking radiation, and possibly evaporate so that the box is filled with a charged gas of electrons or other particles.  More work is needed to understand charged thermal gasses in this context.  One direction for future research is to add charged fields to the action and determine how they affect the thermodynamics.  It seems that fixing the potential at the boundary (grand canonical enesmble) rather than the charge inside makes more physical sense when the black hole can emit charged particles, but the canonical ensemble may still have a role to play.

\begin{acknowledgements}
The author would like to thank James York for many helpful discussions.
\end{acknowledgements}

\end{document}